\documentclass[%
 reprint,
 amsmath,amssymb,
 aps,superscriptaddress,
]{revtex4-2}

\usepackage{graphicx}
\usepackage{dcolumn}
\usepackage{bm}
\usepackage{braket}
\usepackage{xfrac}
\usepackage[separate-uncertainty = true, multi-part-units=single]{siunitx}
\usepackage{xcolor}
\usepackage{gensymb}
\usepackage[english]{babel}
\usepackage{natbib}
\usepackage{relsize}

\newcommand{\CNOT}[0]{{\small CNOT}}
\newcommand{\QWCNOT}[0]{{\small QW-CNOT}}

\newcommand{\QW}[0]{{\small QW}}
\newcommand{\SPDC}[0]{{\small SPDC}}
\newcommand{\DC}[0]{{\small DC}}
\newcommand{\LNOI}[0]{{\small LNOI}}
\newcommand{\HOM}[0]{{\small HOM}}
\newcommand{\CHSH}[0]{{\small CHSH}}

\newcommand{\FidelityClassical}[0]{\SI{0.982(7)}{}}

\newcommand{\FidelityTransferIdealMixed}[0]{\SI{0.994(1)}{}}
\newcommand{\FidelityTransferIdealPure}[0]{\SI{0.938(3)}{}}
\newcommand{\FidelityTransferSimMixed}[0]{\SI{0.991(1)}{}}
\newcommand{\FidelityTransferSimPure}[0]{\SI{0.985(2)}{}}

\newcommand{\FidelityBellIdealMixed}[0]{\SI{0.985(1)}{}}
\newcommand{\FidelityBellIdealPure}[0]{\SI{0.945(2)}{}}
\newcommand{\FidelityBellSimMixed}[0]{\SI{0.987(1)}{}}
\newcommand{\FidelityBellSimPure}[0]{\SI{0.973(1)}{}}

\begin{document}

\preprint{APS/123-QED}

\title{Quantum logical controlled-NOT gate in a lithium niobate-on-insulator photonic quantum walk}

\author{Robert J. Chapman}
\email{rchapman@ethz.ch}
\affiliation{Optical Nanomaterial Group, Institute for Quantum Electronics, Department of Physics, ETH Zurich, CH-8093 Zurich, Switzerland}

\author{Samuel H\"ausler}
\affiliation{Institute for Sensors and Electronics, University of Applied Sciences and Arts Northwestern Switzerland, CH-5210 Windisch, Switzerland}

\author{Giovanni Finco}
\affiliation{Optical Nanomaterial Group, Institute for Quantum Electronics, Department of Physics, ETH Zurich, CH-8093 Zurich, Switzerland}

\author{Fabian Kaufmann}
\affiliation{Optical Nanomaterial Group, Institute for Quantum Electronics, Department of Physics, ETH Zurich, CH-8093 Zurich, Switzerland}

\author{Rachel Grange}
\affiliation{Optical Nanomaterial Group, Institute for Quantum Electronics, Department of Physics, ETH Zurich, CH-8093 Zurich, Switzerland}

\date{\today}

\begin{abstract}
Quantum computers comprise elementary logic gates that initialize, control and measure delicate quantum states.
One of the most important gates is the controlled-NOT, which is widely used to prepare two-qubit entangled states.
The controlled-NOT gate for single photon qubits is normally realized as a six-mode network of individual beamsplitters.
This architecture however, utilizes only a small fraction of the circuit for the quantum operation with the majority of the footprint dedicated to routing waveguides.
Quantum walks are an alternative photonics platform that use arrays of coupled waveguides with a continuous interaction region instead of discrete gates.
While quantum walks have been successful for investigating condensed matter physics, applying the multi-mode interference for logical quantum operations is yet to be shown.
Here, we experimentally demonstrate a two-qubit controlled-NOT gate in an array of lithium niobate-on-insulator waveguides.
We engineer the tight-binding Hamiltonian of the six evanescently-coupled single-mode waveguides such that the multi-mode interference corresponds to the linear optical controlled-NOT unitary.
We measure the two-qubit transfer matrix with \FidelityTransferIdealPure{} fidelity, and we use the gate to generate entangled qubits with \FidelityBellIdealPure{} fidelity by preparing the control photon in a superposition state.
Our results highlight a new application for quantum walks that use a compact multi-mode interaction region to realize large multi-component quantum circuits.
\end{abstract}

\maketitle

\begin{figure*}
    \centering
  \includegraphics[width=1\linewidth,trim={30 20 30 20},clip]{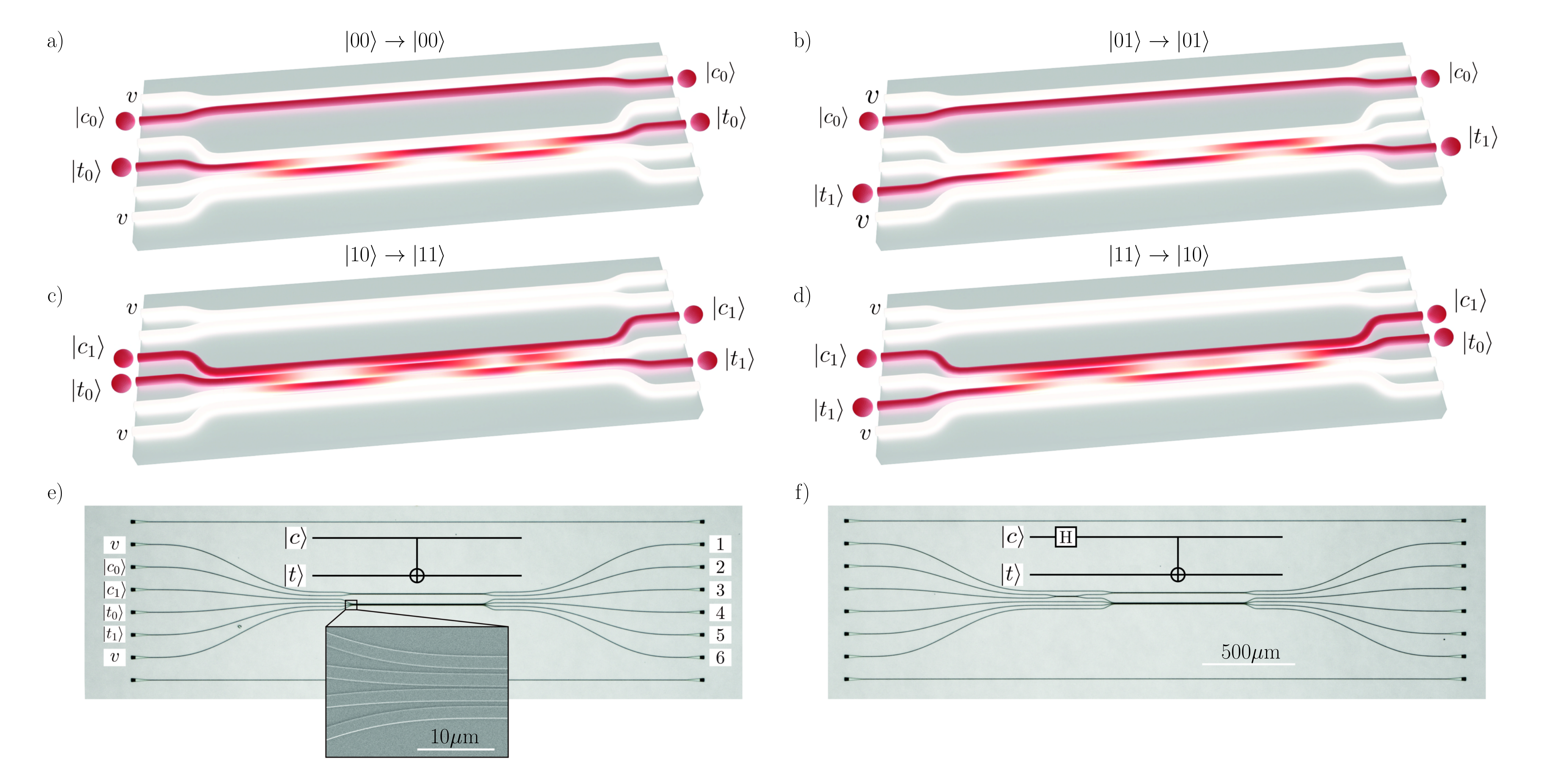}
\caption{\textbf{Evolution of the \QWCNOT{} device.} 
Two-qubit evolution in the \QWCNOT{} waveguide array shown in the two-qubit logical subspace for input states $\ket{ct}$: a) $\ket{00}$, b) $\ket{01}$, c) $\ket{10}$, and d) $\ket{11}$. 
Labels $\ket{c}$ and $\ket{t}$ indicate the control and target qubits, and the input waveguides labeled $v$ are auxiliary modes.
When the control qubit is in the $\ket{1}$ state the photon interferes with the target qubit and causes a bit-flip operation. 
This operation is only possible when post-selecting the logical subspace.
e) A microscope image of the \QWCNOT{} realized in the \LNOI{} photonics platform. The labels on the left show the qubit encoding and on the right the waveguide number. The inset shows a scanning electron micrograph of the start of the coupling region.
f) A microscope image of the \QWCNOT{} array with an additional \DC{} that implements a Hadamard gate, labeled H, preparing the control qubit in a superposition state.
}
\label{fig:fig0}
\end{figure*}

Photonic quantum information technologies use controlled superposition and entangled single photon qubits for applications in quantum computing \cite{obrien_optical_2007, slussarenko_photonic_2019}, simulation \cite{aspuru-guzik_photonic_2012}, communication \cite{gisin_quantum_2007} and sensing \cite{pirandola_advances_2018}.
At the core of any quantum information processor is the ability to entangle qubits via logical operation such as the controlled-NOT (\CNOT{}) gate, which is fundamental to many leading quantum computing algorithms.
These operations are typically the most challenging to perform, as the qubits must interact with each other but be otherwise fully isolated to preserve coherence \cite{ladd_quantum_2010}.
In linear optical quantum computing, photonic qubits are entangled by quantum interference at beamsplitters and single photon detection, either in post-selection or by heralding ancillary photons \cite{knill_scheme_2001}.
The typical linear optical \CNOT{} gate is realized in a six-mode interferometer comprising five beamsplitters \cite{ralph_linear_2002} and has been demonstrated in free-space optics with polarization encoding \cite{obrien_demonstration_2003}, and in integrated photonics with path encoding \cite{politi_silica--silicon_2008}.
Four modes are for the two qubits and the extra modes provide loss channels that are necessary to balance the probabilistic logical operation.
The need for high-fidelity entangling gates puts stringent requirements on the accuracy of the beamsplitter reflectivities, stability of the interferometer and indistinguishability of the single photons in all degrees of freedom.
Photonic integrated circuits benefit from the precision and inherent phase stability of monolithic nanofabricated devices, and comprise optical waveguides for routing light on-chip, directional couplers (\DC{}s) or multi-mode interferometers that act as two-mode beamsplitters, and thermo- or electro-optic phase control for reconfigurability.
It is known that a nest of two-mode interferometers can be configured for any linear optical unitary, including the \CNOT{} gate, however, this 
architecture is dominated by routing of light between separate beamsplitters, which does not contribute to the logical operation \cite{reck_experimental_1994, clements_optimal_2016}.
 
Quantum walks (\QW{}s) are an alternative photonics architecture, realized as arrays of coupled waveguides where light interferes along the entire propagation length instead of at individual beamsplitters.
The optical evolution in the \QW{} is described by a tight-binding Hamiltonian that can be engineered for many applications and for studying fundamental physics.
This includes the experimental demonstration of topologically bound states \cite{kraus_topological_2012, hafezi_imaging_2013, rechtsman_photonic_2013, mittal_topological_2018, zilberberg_photonic_2018, tambasco_quantum_2018}, Anderson localization \cite{schwartz_transport_2007, lahini_anderson_2008, martin_anderson_2011, crespi_anderson_2013, blanco-redondo_topological_2018}, quantum transport \cite{perez-leija_coherent_2013, chapman_experimental_2016} and for preparing large single photon superposition states \cite{grafe_-chip_2014}.
When pairs of indistinguishable photons propagate in a \QW, they undergo bosonic bunching similar to the Hong-Ou-Mandel (\HOM{}) effect, which has been observed in free-space optics \cite{schreiber_photons_2010}, multi-mode fiber-optics \cite{defienne_two-photon_2016} and integrated photonics \cite{peruzzo_quantum_2010}.
It has been recently proposed that multi-mode quantum interference in a specifically parameterized six-waveguide array can implement the \CNOT{} gate on path-encoded photonic qubits \cite{lahini_quantum_2018}.
In this realization of the \CNOT{} gate, the photons interfere continuously along the length of the array, removing the need for on-chip routing.
However, such a device is challenging to produce because of the precise control required on the tight-binding Hamiltonian for each propagation and hopping term.
While sophisticated \QW{}s have been demonstrated in several photonics technologies, controlled two-qubit gates are yet to be realized.

Here, we experimentally demonstrate the two-qubit \CNOT{} gate realized in a continuous time \QW{}.
We fabricated the six-waveguide ``\QWCNOT{}'' array in lithium niobate-on-insulator (\LNOI{}) and fully control the on-site energies and nearest-neighbor hopping of the tight-binding Hamiltonian in the design of the individual waveguide widths and separations respectively.
We measure the two-qubit \CNOT{} transfer matrix with a fidelity of \FidelityTransferIdealPure{} using photons generated by spontaneous parametric down-conversion (\SPDC{}).
The reduced fidelity is due to the limited indistinguishability of the \SPDC{} source, bandwidth of the \QWCNOT{} chip and fiber dispersion.
We also prepare the control qubit in the $\tfrac{1}{\sqrt{2}}(\ket{0} + e^{i\phi}\ket{1})$ superposition state using an on-chip directional coupler and use the \QWCNOT{} gate to generate the state $\tfrac{1}{\sqrt{2}}(\ket{00} + e^{i\phi}\ket{11})$ with fidelity \FidelityBellIdealPure{}, measured in the computational basis.
The addition of on-chip phase control would enable generation of maximally entangled Bell states, which are an important resource for quantum computing and quantum communication.
Our results open a pathway towards implementing large multi-mode photonic circuits in a single step using \QW{}s.

\section*{Results}

\subsection*{Quantum controlled-NOT gate in a photonic quantum walk}

The quantum walk controlled-NOT (\QWCNOT{}) gate operates on path encoded control $\ket{c}$ and target $\ket{t}$ qubits, that are superposition states of photon occupation across pairs of neighboring waveguides.
The evolution of the quantum walk is described by the tight-binding Hamiltonian
\begin{equation}
    h = \sum_{i=1}^N \beta_i \hat{a}_i^\dagger \hat{a}_i + 
    \sum_{i=1}^{N-1} \kappa_i \left( \hat{a}_i^\dagger \hat{a}_{i+1} + \hat{a}_{i+1}^\dagger \hat{a}_i \right),
    \label{eqn:qw_hamiltonian}
\end{equation}
which is the sum of the propagation coefficients of each waveguide $\beta_i = 2\pi n_{\text{eff}, i}/\lambda$ and the coupling rates between neighboring waveguides $\kappa_i$.
The unitary operation is described by Schr\"{o}dinger's equation $U=\exp{(-iht)}$ for an evolution time $t$.
Lahini \textit{et al.} determined the specific six waveguide tight-binding Hamiltonian with an equivalent unitary to the traditional linear optical \CNOT{} gate \cite{lahini_quantum_2018}, which is usually realized as a network of five beamsplitters \cite{ralph_linear_2002} (shown in Supplementary Section \ref{methods:cnot}).
The \QWCNOT{} requires the Hamiltonian-time product
\begin{equation}
    ht=\pi\begin{pmatrix}
        0 & -1.27 & 0 & 0 & 0 & 0 \\
        -1.27 & -0.73 & 0 & 0 & 0 & 0 \\
        0 & 0 & 0.67 & -0.51 & 0 & 0 \\
        0 & 0 & -0.51 & 0.01 & -1.69 & 0 \\
        0 & 0 & 0 & -1.69 & -1.01 & -0.52 \\
        0 & 0 & 0 & 0 & -0.52 & -1.67 
    \end{pmatrix},
    \label{eqn:cnot_hamiltonian}
\end{equation}
where diagonal terms correspond to the propagation coefficients $\beta_i$ and the off-diagonal terms correspond to the nearest-neighbor hopping rates $\kappa_i$.
The two-photon evolution for each input state is shown in Figure \ref{fig:fig0} for the reduced logical subspace.
Four waveguides are required to encode the two qubits and two additional auxiliary modes that allow the photons to propagate in a larger Hilbert space that is subsequently reduced by post-selection.
Figures \ref{fig:fig0}a and \ref{fig:fig0}b show the control qubit in the $\ket{0}$ state where it is decoupled from the rest of the array and does not interfere with the target qubit.
Coupling to the auxiliary modes and to other non-logical modes are not shown in this evolution, but can be seen in the complete Hilbert space (shown in Supplementary Section \ref{methods:design}).
Figures \ref{fig:fig0}c and \ref{fig:fig0}d show the control qubit in the $\ket{1}$ state, where it quantum interferes with the target qubit.
Quantum interference between the two photons leads to a bit-flip operation on the target qubit.
The visualization here does not consider the amplitude of the state within the two-qubit subspace where each of the logical transformations has a $1/9$ success probability (shown in Supplementary Section \ref{methods:design}).

\begin{figure}
    \centering
  \includegraphics[width=1\linewidth,trim={10 10 10 10},clip]{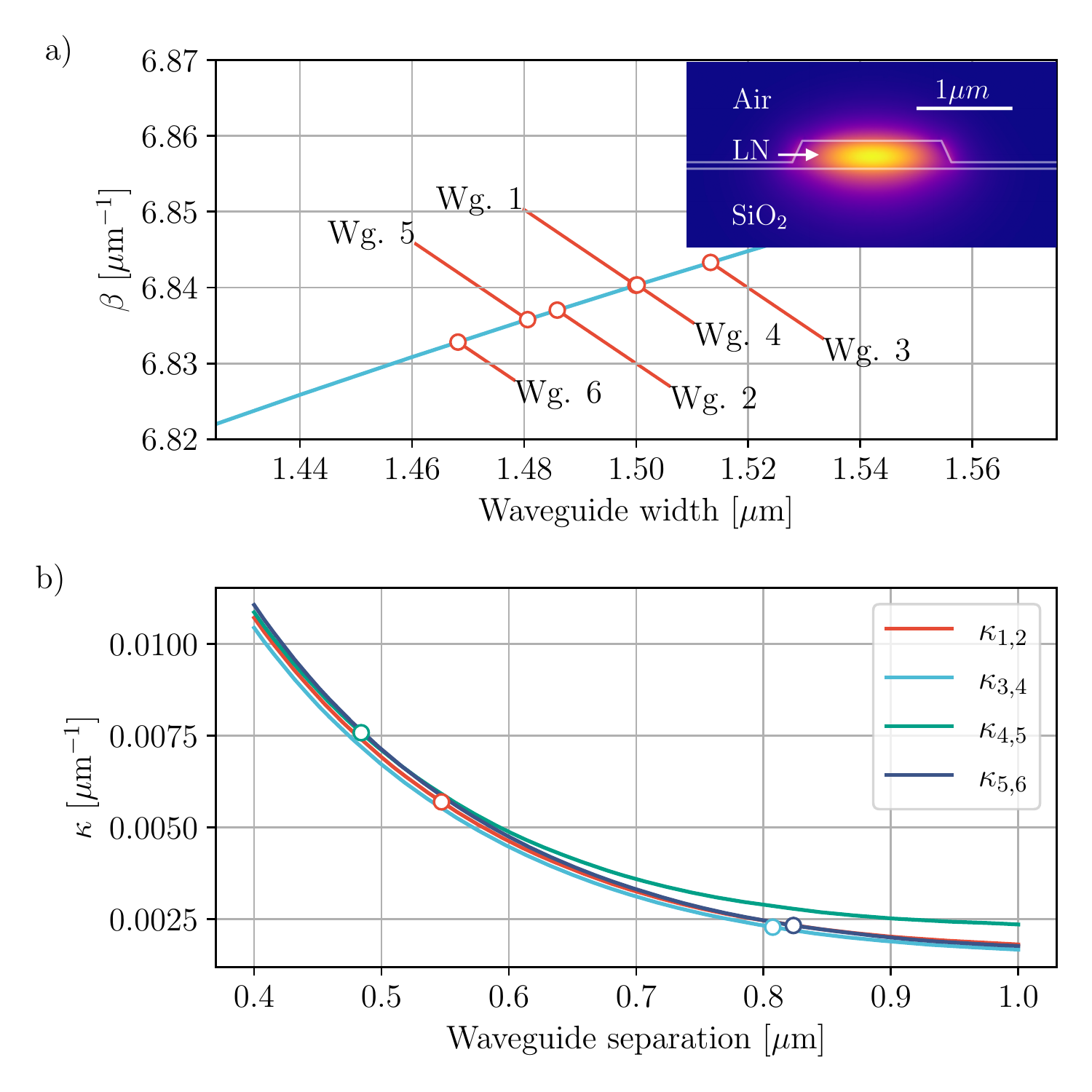}
\caption{\textbf{Design of the \QWCNOT{} array in \LNOI{} photonics.} a) The dependence of the propagation coefficient $\beta$ on the waveguide width (top edge) with a fixed etch depth of \SI{230}{nm} and remaining film of \SI{70}{nm}. The required waveguide widths are indicated on the curve. The inset shows the fundamental transverse electric mode of the \LNOI{} waveguide. b) The dependence of the coupling coefficient $\kappa$ on the waveguide separation (edge to edge) for each of the coupled waveguides. Waveguides 2 and 3 are not coupled as $\kappa_2=0$ in the Hamiltonian.}
\label{fig:fig1}
\end{figure}

\begin{figure*}
    \centering
  \includegraphics[width=1\linewidth,trim={100 0 100 20},clip]{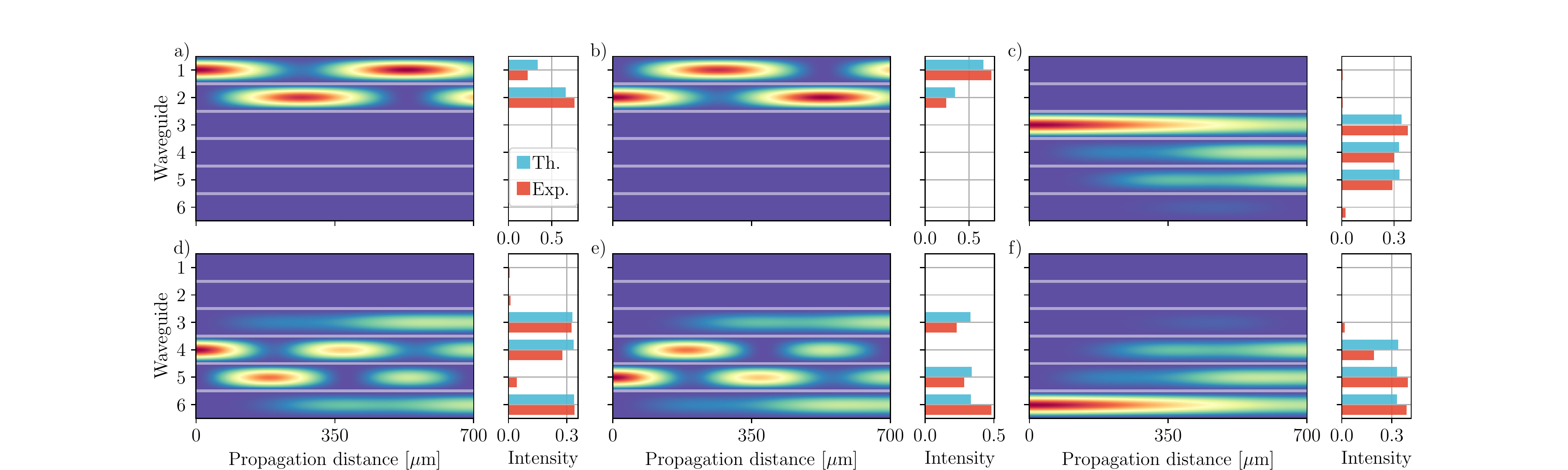}
\caption{\textbf{Classical characterization.} The theoretical evolution intensity in the \QWCNOT{} chip for each of the six waveguide inputs. The output intensity for each evolution is shown for the theoretical model (blue) and the experimental result (red). The theoretical evolution is calculated from the Hamiltonian in Equation \ref{eqn:qw_hamiltonian} with the values of $\beta$ and $\kappa$ shown in Figure \ref{fig:fig1}.}
\label{fig:fig2}
\end{figure*}

\subsection*{\QWCNOT{} design and fabrication}

We use the lithium niobate-on-insulator (\LNOI) photonics platform to implement a \QWCNOT{} gate with waveguides designed for \SI{1550}{\nm} wavelength and transverse electric polarization. 
\LNOI{} has a broad transparency range, low loss, and high electro-optic and $\chi^{(2)}$ nonlinear coefficients, making it a leading technology for photonic quantum information processors \cite{zhu_integrated_2021}.
\LNOI{} has become a major focus in photonics research, with demonstrations of record high-speed and low-voltage modulation \cite{wang_integrated_2018}, ultra-efficient nonlinear frequency conversion \cite{lu_periodically_2019}, all-optical switching \cite{guo_femtojoule_2022} and entangled photon pair generation \cite{zhao_high_2020}.
We etch the waveguide array in a \SI{300}{\nm} LN film with a \SI{4.7}{\um} SiO$_2$ bottom oxide on a silicon handle wafer \cite{kaufmann_redeposition-free_2023}.
The etched waveguide height is \SI{230}{\nm} and we use the width and separation to control the propagation coefficients $\beta_i$ and coupling rates $\kappa_i$ in the tight-binding Hamiltonian.
We initially simulate the mode for various waveguide widths using a finite-element solver and plot the corresponding $\beta$ terms in Figure \ref{fig:fig1}a.
The inset in Figure \ref{fig:fig1}a shows the typical waveguide cross section and solution of the fundamental transverse electric mode.
We exchange time-evolution for equivalent length-evolution for designing the array and target an array length $L=\SI{700}{\um}$ such that the Hamiltonian-length product $hL$ matches Equation \ref{eqn:cnot_hamiltonian}.
We chose the on-site terms to be relative about the first waveguide which we set to \SI{1.5}{\um} width and indicate the necessary propagation coefficients as points in Figure \ref{fig:fig1}a.
Using the determined waveguide widths, we simulate the even and odd supermodes of neighboring waveguides to calculate the coupling rate as $\kappa=\pi(n_{even}-n_{odd})/\lambda$, as shown in Figure \ref{fig:fig1}b. 
The points indicate the necessary parameters for the Hamiltonian.
Note that there is no hopping between waveguides $2$ and $3$ in the Hamiltonian as $\kappa_2=0$.
The waveguide numbering is shown in Figure \ref{fig:fig0}e, and we position waveguides $2$ and $3$ with a \SI{20}{\um} gap to effectively decouple the modes.
Figures \ref{fig:fig0}e shows a microscope image of the \QWCNOT{} and Figure \ref{fig:fig0}f shows another device on the same chip with an additional \DC{} that prepares the control qubit in a superposition state.
The different waveguide separations can be seen in the scanning electron micrograph and reflects the different coupling coefficients of the Hamiltonian.
We couple light to the chip using grating couplers with \SI{\sim6}{dB} loss per coupler. 
Grating couplers offer several advantages over end-fire coupling, such as eliminating the need for dicing and polishing waveguide facets, larger mode field diameters, and polarization sensitivity, ensuring the excitation of only the waveguide transverse electric mode.
Higher efficiency grating couplers in \LNOI{} have been reported \cite{krasnokutska_high_2019, kang_high-efficiency_2020, lomonte_efficient_2021} with losses as low as \SI{<1}{dB} per grating \cite{chen_low-loss_2022}.

\subsection*{Classical characterization}

To classically characterize the \QWCNOT{} gate, we inject a \SI{1550}{\nm} wavelength continuous-wave laser into each input of the chip and record the intensity at all the outputs with a near-infrared camera.
Figure \ref{fig:fig2} shows the propagation simulation based on the Hamiltonian in Equation \ref{eqn:cnot_hamiltonian}, as well as measured output intensity distributions for each of the six inputs in the waveguide array.
The propagation here differs from the two-photon evolution shown in Figure \ref{fig:fig0} as only a single mode is excited.
We calculate the fidelity between the experimental transfer matrix $\Gamma$ and the theoretical transfer matrix $\Gamma'$ as
\begin{equation}
    F(\Gamma, \Gamma') = \mathlarger{\mathlarger{\sum}}_{i,j}\sqrt{\frac{\Gamma_{i,j}\Gamma'_{i,j}}{\left(\sum_{i,j}\Gamma_{i,j}\right)\left(\sum_{i,j}\Gamma'_{i,j}\right)}}.
    \label{eqn:fidelity}
\end{equation}
The fidelity between the classically characterized transfer matrix and the theoretical model is \FidelityClassical{} where the error is calculated from the variation in fidelity for each waveguide input. 
From this characterization, we can build a simulation of the device by finding the Hamiltonian that gives the closest match to the measured output. 
This simulation will be a tool for predicting the two-photon dynamics of the \QWCNOT{} chip.

\begin{figure*}
    \centering
  \includegraphics[width=1\linewidth,trim={0 0 0 0},clip]{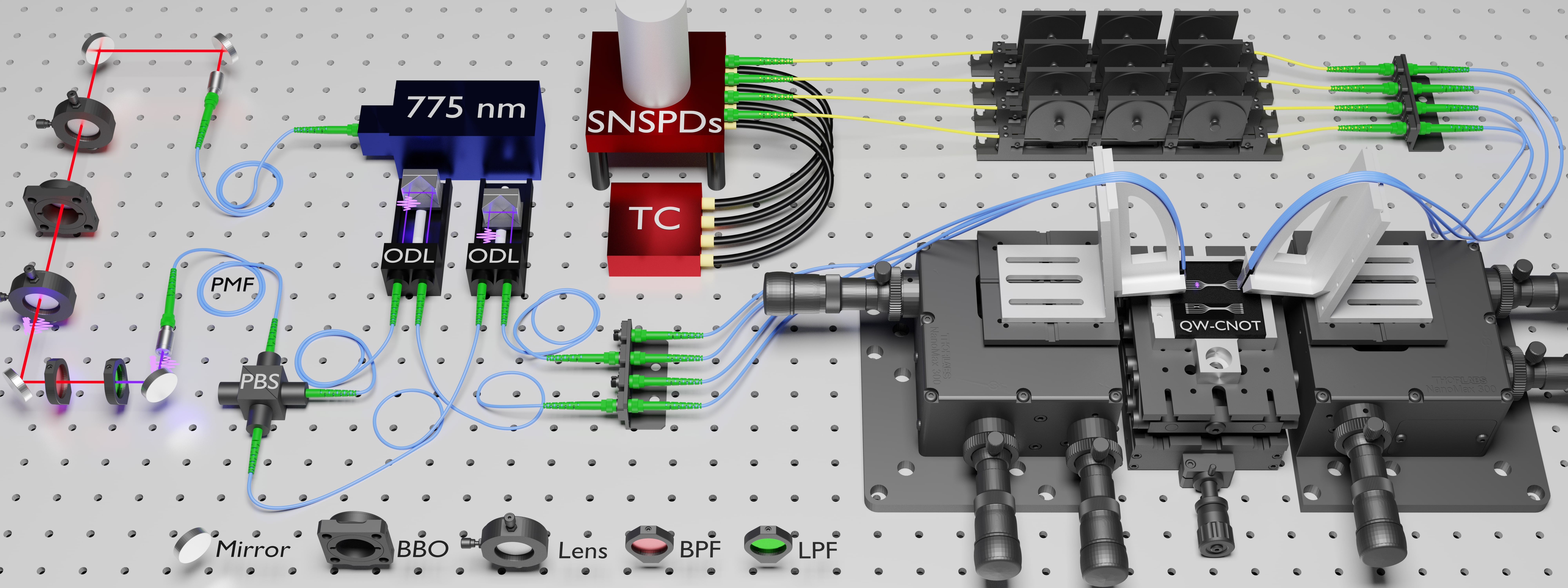}
\caption{\textbf{Experimental setup.} Orthogonally polarized photon-pairs are generated by type-2 spontaneous parametric down-conversion in a \SI{2}{mm}-thick beta barium borate (BBO) crystal and the pump is filtered with a longpass filter (LPF) and the photons with a \SI{12}{nm} bandpass filter (BPF). The photons are coupled to polarization maintaining fiber (PMF) and separated with a polarizing beamsplitter (PBS) where one fiber pigtail is rotated such that both photons have the same output polarization. The photons are injected into optical delay lines (ODLs) to compensate for fiber length mismatch and fiber birefringence and are coupled to the \QWCNOT{} chip with a fiber array positioned above on-chip grating couplers. After the quantum walk, the photons are collected in fiber and measured with superconducting nanowire single photon detectors (SNSPDs) and the arrival times are recorded with a time correlator (TC). Polarization controllers are used to maximize the efficiency of the SNSPDs.}
\label{fig:fig3}
\end{figure*}

\subsection*{\SPDC{} source and experimental setup}

In order to measure the two-qubit operation of the \QWCNOT{} gate, we require a source of indistinguishable photons at \SI{1550}{\nm} wavelength. 
To achieve this, we employ a collinear type-2 spontaneous parametric down-conversion (\SPDC{}) process using a beta-barium borate crystal. 
A diagram of the complete experimental setup is depicted in Figure \ref{fig:fig3}. 
We use a \SI{775}{\nm} continuous-wave laser to pump the nonlinear crystal, generating photon pairs which are filtered with a \SI{12}{\nm} bandpass filter to ensure both photons have the same bandwidth. 
Because we use a type-2 \SPDC{} process, one photon is horizontally polarized and the other vertically polarized. 
We couple both photons into a polarization maintaining fiber and separate them with a fiber-pigtailed polarizing beamsplitter.
One of the output fibers is pigtailed with the slow axis vertical and the other fiber with the slow axis horizontal, so that both photons exit the beamsplitter with the same polarization in each fiber.
Motor-controlled optical delay lines compensate for mismatch in fiber lengths and birefringence before the photons are coupled to the chip.
Photons are coupled to and from the chip with fiber arrays positioned above the grating couplers with precision translation stages.
Finally, coincidence measurements are performed with superconducting nanowire single photon detectors and a time correlator to record arrival times.
We verify the indistinguishability of the generated photons through the characteristic \HOM{} dip, measured by interfering the photons in a fused-silica 50:50 beamsplitter with \SI{94.6}{\%} visibility (further details in Supplementary Section \ref{methods:SPDC}).
Narrower bandpass filters can improve the source visibility by reducing fiber dispersion at the cost of lowering the photon count rate.

\subsection*{\QWCNOT{} array two-qubit transfer matrix}

\begin{figure*} 
    \centering
  \includegraphics[width=1\linewidth,trim={125 80 60 100},clip]{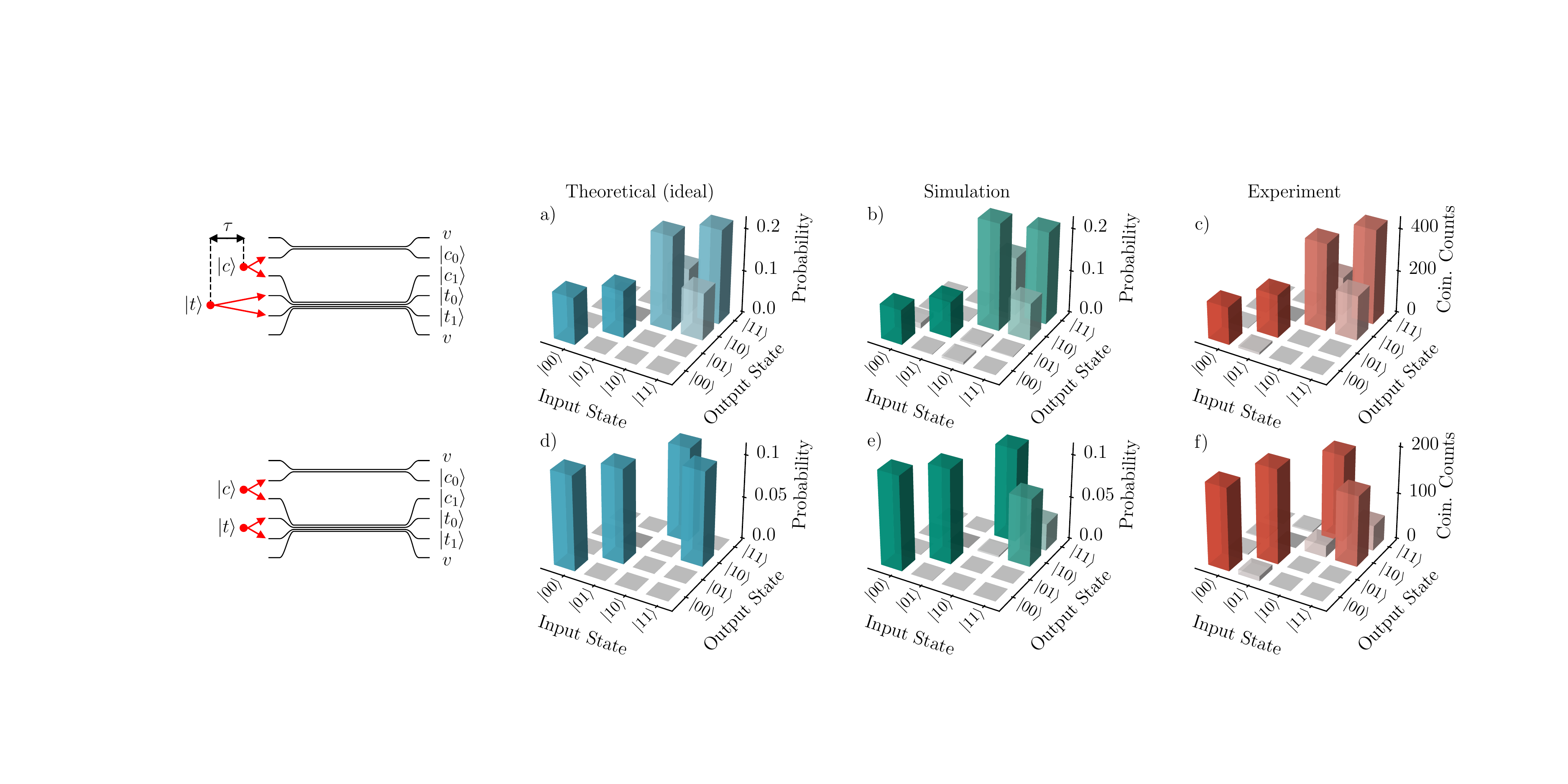}
\caption{\textbf{Two-qubit transfer matrix.} 
a, b, c) The theoretical (ideal), simulated and experimental two-qubit transfer matrix when the photons arrive with a time separation $\tau$. In this regime, the photons do not quantum interfere. d, e, f) The theoretical (ideal), simulated and experimental two-qubit transfer matrix with indistinguishable photons that arrive with zero time delay. 
}
\label{fig:fig4}
\end{figure*}

The operation of the two-qubit \QWCNOT{} chip is based on quantum interference between the control and target qubits. 
The control $\ket{0}$ state does not interfere with the target qubit and thus the arrival time of the two photons has no impact on the transfer matrix. 
However, for the control $\ket{1}$ state, the photon arrival times are critical to allow for interference. 
We use the motorized optical delay lines to match the arrival times of the control and target qubits. 
For the input state $\ket{10}$, we observe \HOM{} interference between waveguides $3$ and $4$, which results in suppression of the output state $\ket{10}$ and therefore only the logical output $\ket{11}$ is measured. 
Likewise for input state $\ket{11}$, \HOM{} interference is observed between waveguides $3$ and $5$, meaning only the logical output $\ket{10}$ is measured.
This interference gives rise to the logical \CNOT{} operation in our chip.
The raw \HOM{} interference plots are shown in Supplementary Section \ref{methods:two-photon}.
We measure the two-qubit transfer matrix of the \QWCNOT{} chip with and without a delay between the photons and plot the results in Figure \ref{fig:fig4}.
When the photons arrive at the chip with a time delay ($\tau$), there is no quantum interference and we observe classical propagation dynamics.
In Figures \ref{fig:fig4}a, \ref{fig:fig4}b and \ref{fig:fig4}c, we show the ideal theoretical, simulation and experimental results respectively.
Because there is no quantum interference, the simulation is simply the product of the classical probability distributions shown in Figure \ref{fig:fig2}.
We calculate the fidelity using Equation \ref{eqn:fidelity} between the theoretical model and the experiment as \FidelityTransferIdealMixed{} and between the simulation and the experiment as \FidelityTransferSimMixed{}, where errors are calculated using Poissonian statistics.

We scan the optical delay lines and measure \HOM{} interference between the control and target photons in the \QWCNOT{} waveguide array.
When the delay is $\tau=0$, the photons arrive simultaneously and the \CNOT{} operation is implemented by quantum interference.
The theoretical, simulated and experimentally measured transfer matrices are shown in Figures \ref{fig:fig4}d, \ref{fig:fig4}e and \ref{fig:fig4}f respectively.
The fidelity between the theoretical model and the experiment is \FidelityTransferIdealPure{} and between the simulation and the experiment is \FidelityTransferSimPure{}.
The reduced fidelity of the two-photon interference is attributed to the relatively broad bandwidth of the \SPDC{} photons (\SI{12}{\nm}), meaning optical dispersion in fibers and waveguide, and the sensitivity of the Hamiltonian to wavelength, decrease the visibility of the two-qubit interference.

\subsection*{Entangled state generation}

\begin{figure*} 
    \centering
  \includegraphics[width=1\linewidth,trim={10 10 10 10},clip]{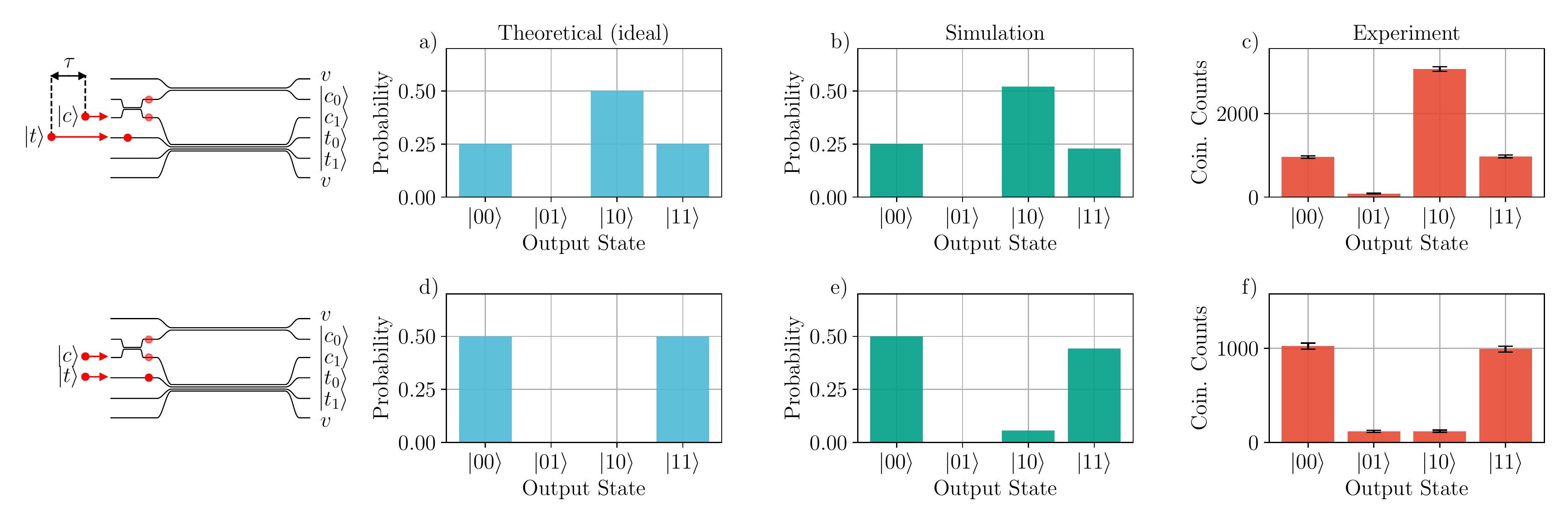}
\caption{\textbf{Entangled state preparation.} 
A directional coupler is included before the \QWCNOT{} array to prepare the control qubit in a superposition state $(\ket{0}+e^{i\phi}\ket{1})/\sqrt{2}$.
a, b, c) The theoretical (ideal), simulated and experimental two-qubit output when the photons arrive with a time difference $\tau$.
This means the two qubits are distinguishable and no entanglement is generated.
d, e, f) The theoretical (ideal), simulated and experimental two-qubit output when the photons arrive together and undergo \HOM{} interference.
In this case, the entangled state $(\ket{00} + e^{i\phi}\ket{11})/\sqrt{2}$ is generated.
The error bars in c) and f) are calculated from Poissonian statistics.
}
\label{fig:fig5}
\end{figure*}

The \CNOT{} gate is a fundamental element for quantum computation and can be used to prepare maximally entangled Bell states when the control qubit is in the superposition state $\ket{c} = \frac{1}{\sqrt{2}}(\ket{0} \pm \ket{1})$ and the target qubit in either the $\ket{0}$ or $\ket{1}$ state.
We fabricated a \QWCNOT{} gate with the addition of a \DC{} that prepares the control qubit in the superposition state $\ket{c} = \frac{1}{\sqrt{2}}(\ket{0} + e^{i\phi}\ket{1})$ and the target qubit is kept in the $\ket{0}$ state.
We do not control the phase $\phi$ in this experiment, however, the addition of electro-optic or thermo-optic phase shifters \cite{maeder_high-bandwidth_2022} would enable the control necessary to prepare maximally entangled Bell states.
When the two photons arrive with a time delay $\tau$, the input state becomes an incoherent mixture $\frac{1}{2}(\ket{00}\bra{00} + \ket{10}\bra{10})$ and no quantum interference is observed. 
The expected output of the \QWCNOT{} in this case is $\frac{1}{4}(\ket{00}\bra{00} + 2\ket{10}\bra{10} + \ket{11}\bra{11})$. 
On the other hand, when the photons arrive simultaneously, i.e $\tau=0$, the input state is a coherent superposition $\frac{1}{\sqrt{2}}(\ket{00} + e^{i\phi}\ket{10})$ and the expected output is the entangled state $\frac{1}{\sqrt{2}}(\ket{00} + e^{i\phi}\ket{11})$. 
The entangled state preparation results are presented in Figure \ref{fig:fig5}. 
Based on the two-photon transfer matrix measured previously, we construct a model of the two-qubit entangled state preparation.
The theoretical, simulated and measured output for the incoherent input state, measured in the computational basis, are shown in Figures \ref{fig:fig5}a, \ref{fig:fig5}b, and \ref{fig:fig5}c respectively.
We calculate the fidelity between the theoretical state and the experiment as \FidelityBellIdealMixed{} and between the simulated state and the experiment as \FidelityBellSimMixed{}.
The theoretical, simulated and measured output for the coherent superposition input state are shown in Figures \ref{fig:fig5}d, \ref{fig:fig5}e, and \ref{fig:fig5}f respectively.
The fidelity between the theoretical output and the experiment is \FidelityBellIdealPure{} and between the simulation and experiment is \FidelityBellSimPure{}.
Although these results are from a different \QWCNOT{} device, the experiment and simulation show good agreement.
The fidelity here assumes $\phi=0$, which can be achieved with an on-chip phase shifter for the input control qubit.
The two-qubit entangled state could be verified by performing two-qubit state tomography or violating the Clauser, Horn, Shimony and Holt (\CHSH{}) inequality \cite{clauser_proposed_1969}. 
Nevertheless, this experiment shows the \QWCNOT{} device is capable of performing photonic entangling gates in a single interaction step, which could provide a significant improvement over the traditional multi-step circuit approach.

\section*{Discussion}

We have successfully implemented the controlled-NOT gate in a continuous-time quantum walk on the lithium niobate-on-insulator photonics platform. 
Using indistinguishable photons from a \SPDC{} source, we have measured the transfer matrix of the two-qubit \CNOT{} gate, and prepared the control qubit in a superposition leading to the generation of entangled states.
Our demonstration does not yet benefit from the key advantages of \LNOI{} photonics, however, we aim to improve gate fidelity by including electro-optical tuning to locally modify the array Hamiltonian.
With enough control, this could also enable us to reconfigure the array for different logical operations and provide a new platform for programmable integrated photonics.
Another improvement would come from adding input and output single qubit gates for performing process and state tomography, violating the \CHSH{} inequality and preparing arbitrary two-qubit states.
In this work, we highlight the potential for photonic quantum gates to be implemented in continuously coupled waveguide arrays, offering a more compact solution compared to existing architectures and opening new pathways towards the development of more complex multi-photon circuits in a single interaction region.

\subsection*{Data availability}

Raw data and evaluation code are available from the authors upon reasonable request.

\subsection*{Competing interests}
The authors declare no competing financial or non-financial interests.

\subsection*{Author Contributions}

R.J.C. conceived the experiment. R.J.C. and S.H designed the waveguide array. G.F. and F.K fabricated the lithium niobate-on-insulator waveguide samples. R.J.C. performed the optical measurements, data analysis and wrote the original draft of the manuscript. R.G. supervised the project. All authors contributed to revising the manuscript.

\subsection*{Acknowledgments}

We acknowledge support for characterization of our samples from the Scientific Center of Optical and Electron Microscopy ScopeM and from the cleanroom facilities BRNC and FIRST of ETH Zurich and IBM Ruschlikon.
R.J.C. acknowledges support from the Swiss National Science Foundation under the Ambizione Fellowship Program (Project Number PZ00P2\_208707).
R.G. acknowledges support from the European Space Agency (Project Number 4000137426), the Swiss National Science Foundation under the Bridge Program (Project Number 194693), the European Research Council (Project Number 714837), and Horizon 2020 (ELENA Consortium, Project Number 101016138).

%

\clearpage
\section*{Supplementary Material}
\renewcommand\thefigure{S\arabic{figure}}    
\setcounter{figure}{0}

\subsection{Linear optical \CNOT{} gate}
\label{methods:cnot}

\begin{figure}
    \centering
    \includegraphics[width=1\linewidth,trim={70 30 70 20},clip]{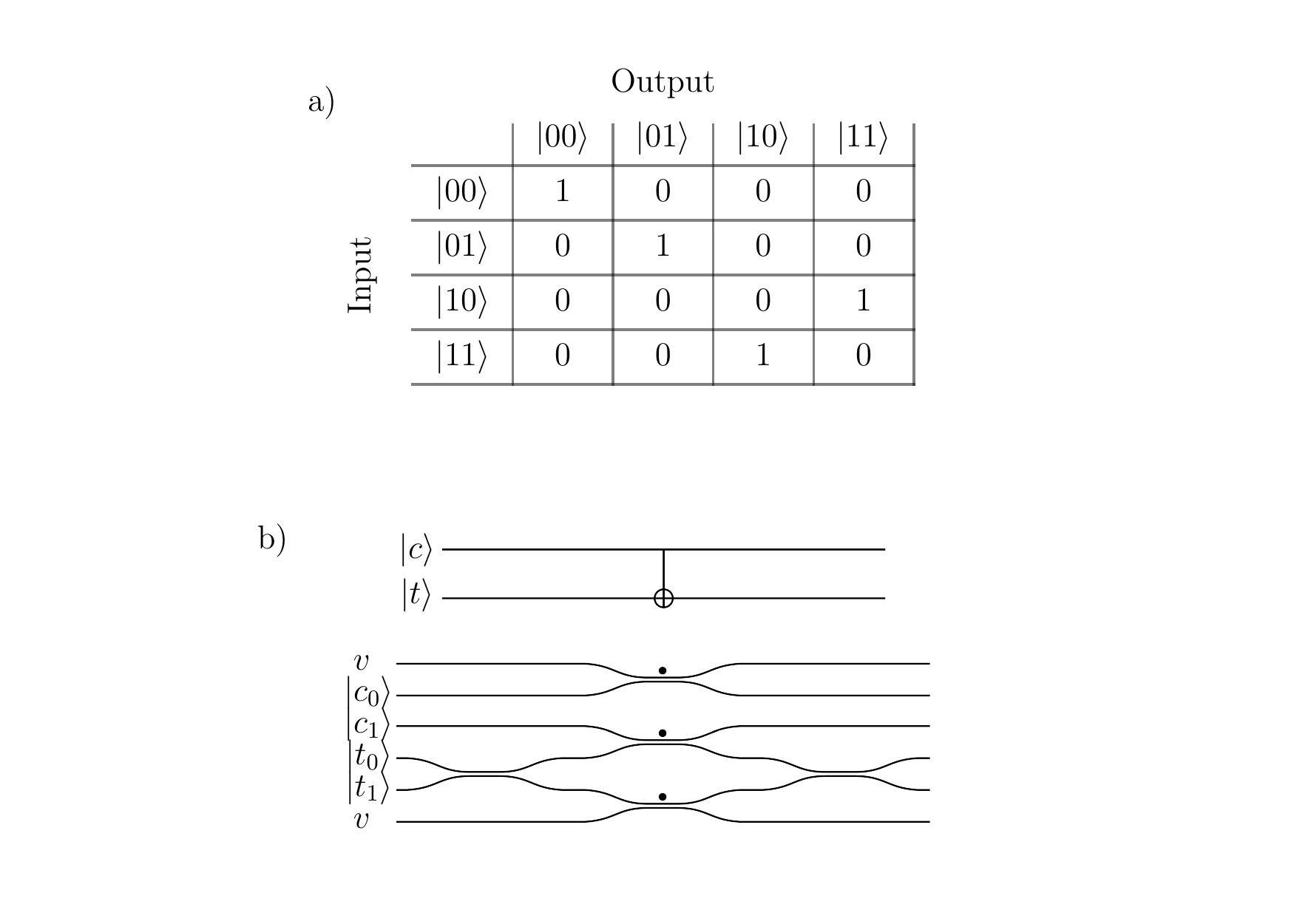}
    \caption{\textbf{Controlled-NOT gate.} a) The truth table of the controlled-NOT gate. Input and output states correspond to $\ket{ct}$ where $\ket{c}$ is the control and $\ket{t}$ is the target qubit. b) Linear optical realization based on $1/2$ and $1/3$ (with a dot) beamsplitters. The gate is post-selected on two photons being in the two-qubit logical subspace. The auxiliary modes $v$ are necessary to balance the post-selected logical operation.}
    \label{fig:si0}
\end{figure}

Figure \ref{fig:si0}a shows the truth table for the controlled-NOT quantum gate.
The state of the target qubit is flipped conditional on the control qubit being in the $\ket{1}$ state.
If the control qubit is in a superposition state, the \CNOT{} gate will prepare an entangled two-qubit state.
Figure \ref{fig:si0}b shows the typical post-selected linear optical \CNOT{} gate.
The two qubits are labeled $\ket{c}$ and $\ket{t}$ for control and target using standard path encoding.
The directional couples without a dot have reflectivity $1/2$ and with a dot have reflectivity $1/3$.
Quantum interference at the central beamsplitter and post-selection lead to generation of entangled states.
The auxiliary modes $v$ are necessary to balance the operation, and the gate has a success probability of $1/9$, where the failed operations exit the two-qubit subspace and are therefore not measured.

\subsection{\CNOT{} array design}
\label{methods:design}

\begin{figure}
    \centering
    \includegraphics[width=1\linewidth,trim={10 10 10 10},clip]{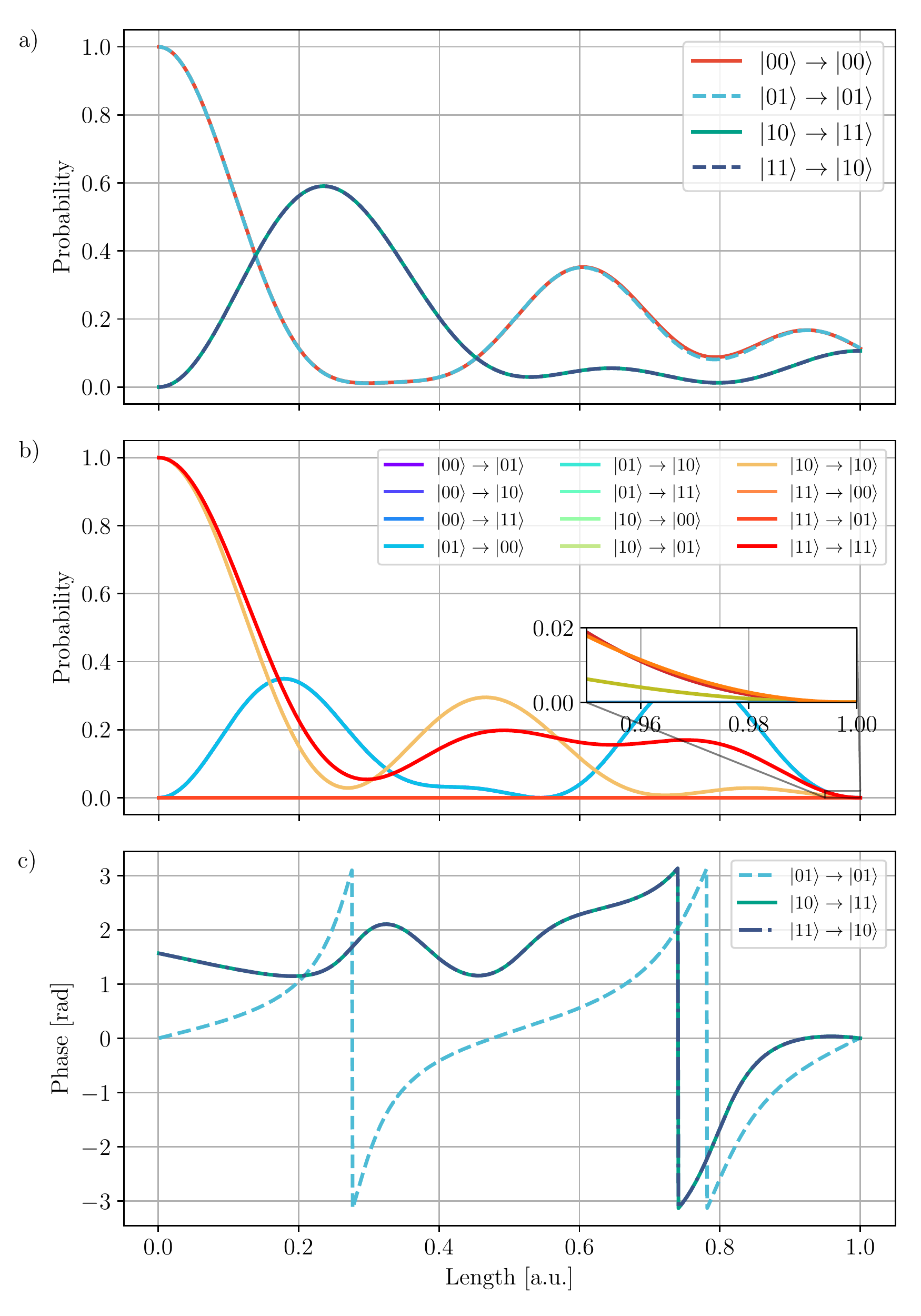}
    \caption{\textbf{\QWCNOT{} evolution.} a) Propagation of the \QWCNOT{} logical operations. All transfers result in $1/9$ probability. b) Propagation of all other trajectories that would lead to logical errors. All trajectories have zero probability at the end of the evolution. c) Relative phase between the logical states involved in the \QWCNOT{} operation. The phase of the $\ket{00}$ is set to zero and all relative phases are zero at the end of the evolution, meaning the operation does not induce any additional phase rotations on the logical qubits. Phases stating at $\pi/2$ is equivalent to the phase shift picked up on reflection.}
    \label{fig:si1}
\end{figure}

\begin{figure*}
    \centering
    \includegraphics[width=1\linewidth,trim={0 100 0 100},clip]{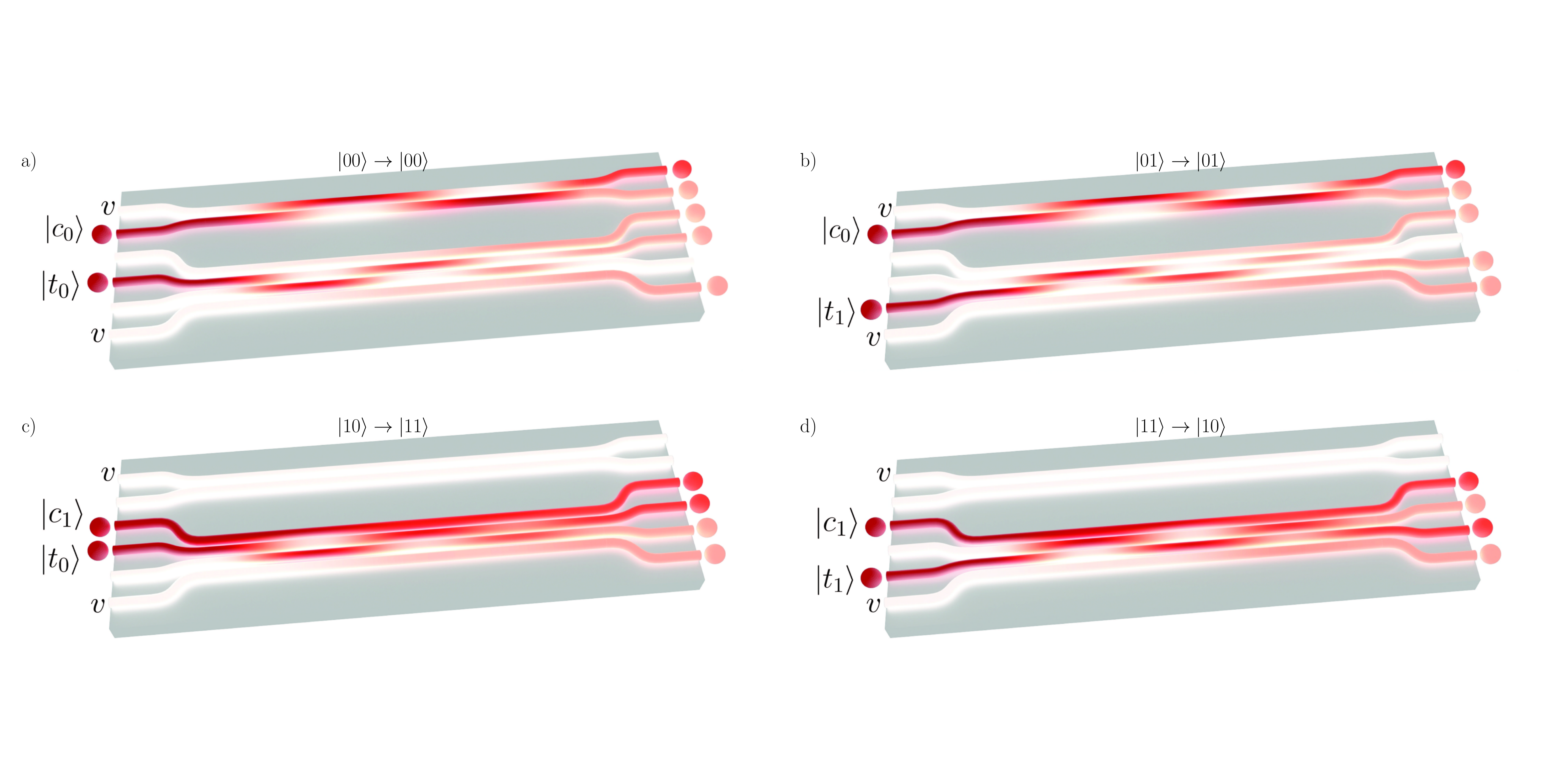}
    \caption{\textbf{\QWCNOT{} gate in the complete Hilbert space.} The trajectories for the two-photon propagation as viewed in the probability distribution of the two photons across the waveguide array. The evolutions include when photons exit the two-qubit subspace. With this plot, it is not possible to see the two-photon correlations that give rise to the \CNOT{} gate.}
    \label{fig:si2}
\end{figure*}

The evolution of the Hamiltonian-time product in Equation \ref{eqn:cnot_hamiltonian} of the main text gives the two-qubit photon \CNOT{} gate operation.
The correct logical transformations have a success probability $1/9$ after evolution and are shown in Figure \ref{fig:si1}a and the other transformations that exit the two-qubit subspace are shown in Figure \ref{fig:si1}b.
The phase of the \CNOT{} transformations are shown Figure \ref{fig:si1}c relative to the phase of the $\ket{00}$ state, demonstrating that the \QWCNOT{} does not add additional phases.
The trajectories of each logical input are shown in Figure \ref{fig:si2} including the complete Hilbert space.

\subsection{SPDC source}
\label{methods:SPDC}

\begin{figure}
    \centering
    \includegraphics[width=1\linewidth,trim={10 10 10 10},clip]{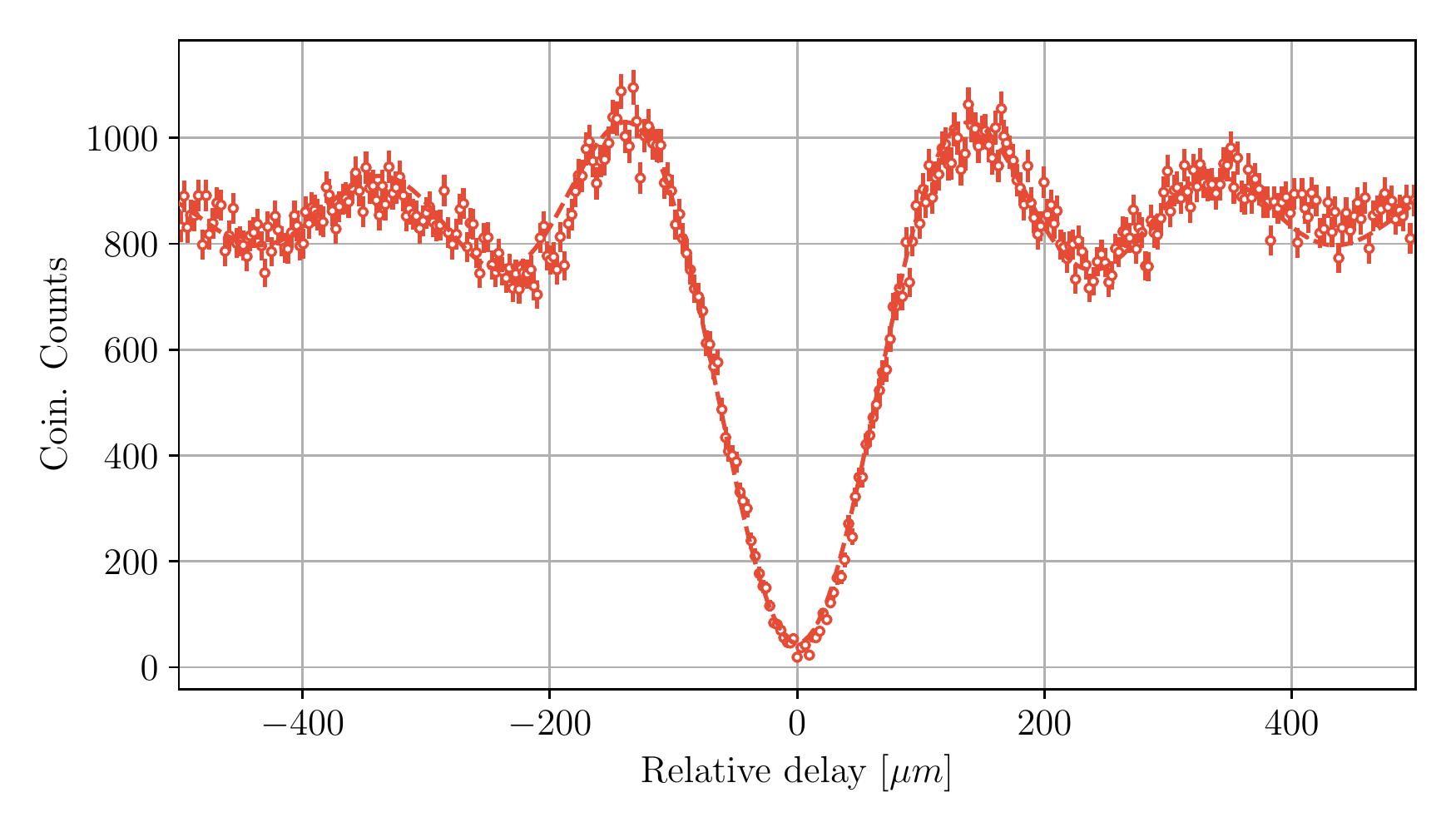}
    \caption{\textbf{SPDC Source Indistinguishability.} Hong-Ou-Mandel interference of the photons generated by SPDC. Error bars are calculated using Poissonian statistics.}
    \label{fig:si3}
\end{figure}

\begin{figure*}
    \centering
    \includegraphics[width=1\linewidth,trim={10 10 90 0},clip]{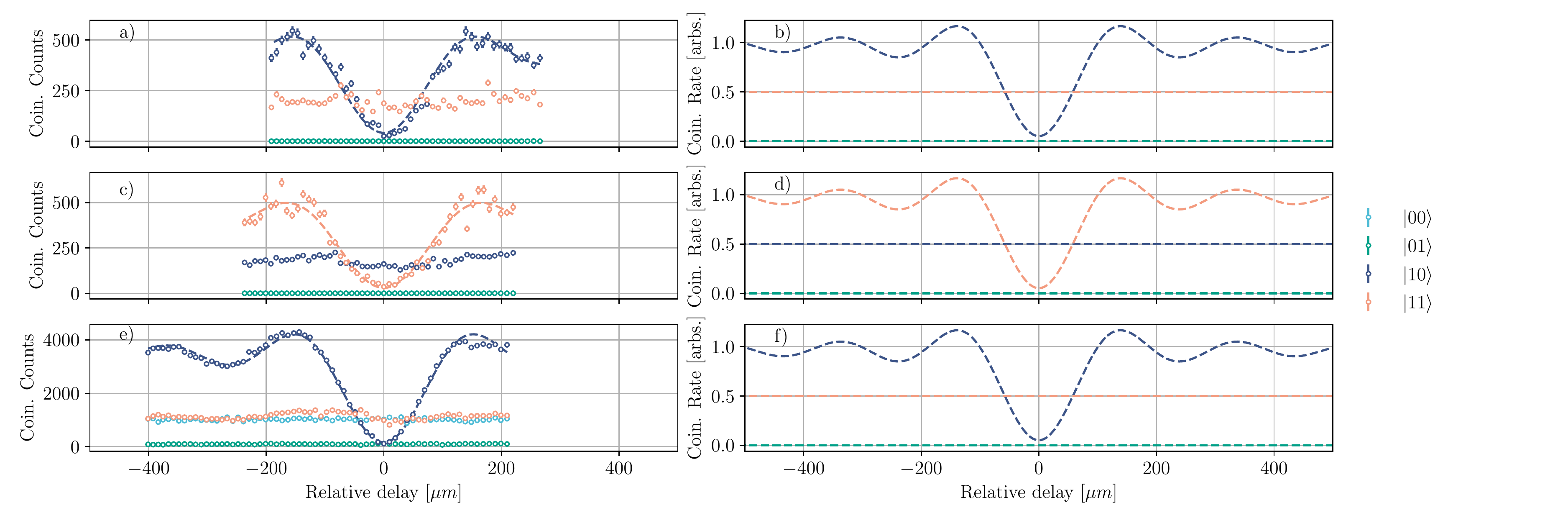}
    \caption{\textbf{Quantum interference in the \QWCNOT.} a, b) Experimental and theoretical values for input state $\ket{10}$. The theoretical value of $\ket{11}$ is constant at $0.5$, $\ket{01}$ quantum interferes and all others are at $0$. c, d) Experimental and theoretical values for input state $\ket{11}$. The theoretical value of $\ket{10}$ is constant at $0.5$, $\ket{11}$ quantum interferes and all others are at $0$. e, f) Experimental and theoretical values for input state $\tfrac{1}{\sqrt{2}}\ket{00}+\ket{10}$. The theoretical value of $\ket{00}$ and $\ket{11}$ are constant at $0.5$, $\ket{10}$ quantum interferes and $\ket{01}$ is at $0$. Error bars are calculated using Poissonian statistics.
    }
    \label{fig:si4}
\end{figure*}

We generate pairs of photons at \SI{1550}{\nm} wavelength in a BBO crystal (Newlight Photonics) pumped with a \SI{\sim350}{mW} \SI{775}{nm} wavelength CW laser (Toptica DL Pro and Newport VAMP). 
The crystal is designed for type-2 collinear SPDC, meaning the photons are generated with orthogonal polarizations.
Then we use a \SI{12}{nm} band pass filter centered at \SI{1550}{\nm} (Thorlabs) to ensure good spectral overlap, and to increase the coherence time of the photons for easier Hong-Ou-Mandel (HOM) interference.
The \SI{12}{nm} filter is also necessary as the Hamiltonian of our device is designed for \SI{1550}{\nm} wavelength and has limited bandwidth.
We couple both photons into a single polarization maintaining fiber (PMF) and separate them using a polarizing beam splitter (PBS, OZ Optics).
Two optical delay lines (ODLs, OZ Optics) allow for \SI{\sim1}{\um} precision in the delay length between the two photons before coupling to the chip using v-groove fiber arrays (OZ Optics) with \SI{8}{\degree} angle from vertical for best grating coupler efficiency.
The input and output fiber arrays with \SI{127}{\um} pitch are mounted on 5 axis translation stages (Thorlabs) for positioning and orienting the gratings.
Alignment waveguides on either side of the array are used to optimize the alignment such that the waveguide array has close to equal coupling efficiency for all modes.
The collected photons are sent to superconducting nanowire single photon detectors (SNSPDs, Single Quantum) with the polarization optimized for detection efficiency.
The electrical signals from the SNSPDs are measured with a time correlator (Swabian Instruments). 
The SNSPDs have \SIrange{15}{50}{ps} jitter and the time correlator \SI{8}{ps} jitter.
A coincidence window of \SI{1}{ns} is used and the coincidence counts between all four logical outputs of the chip are measured simultaneously. 
A typical HOM dip from the source, measured with a fused silica beamsplitter (Thorlabs), is shown in Figure \ref{fig:si3} with visibility \SI{94.6}{\%}.

\subsection{Two photon measurement and analysis}
\label{methods:two-photon}

We measure quantum interference in the \QWCNOT{} chip by scanning one of the ODLs and measuring the coincidence measurements between the four output channels.
We record states that are outside of the logical subspace, however, we do not plot them here for clarity, nor do we measure coincidence in the same output mode via photon number resolving.
For the input states $\ket{00}$ and $\ket{01}$, there is no quantum interference in the chip and we simply measure the product of the output probability distributions.
However, for the input states $\ket{10}$ and $\ket{11}$, the photons undergo quantum interference and therefore the arrival times must match.
Figures \ref{fig:si4}a and \ref{fig:si4}b show the experimental and simulated quantum interference for the input $\ket{10}$ state with \SI{900}{s} integration time for each relative delay.
There is quantum interference between the waveguides $3$ and $4$, which approaches zero when the photons arrive together at the chip.
The simulation here is based on the HOM dip in Figure \ref{fig:si3} to find the temporal length of the photons.
Figures \ref{fig:si4}c and \ref{fig:si4}d show the experimental and simulated quantum interference for the input $\ket{11}$ state, where quantum interference now occurs between waveguides $3$ and $5$.
Figures \ref{fig:si4}e and \ref{fig:si4}f show the experimental and simulated quantum interference when the input state is in a superposition $\tfrac{1}{\sqrt{2}}(\ket{0}+e^{i\phi}\ket{1})$. 
There is again quantum interference between waveguides $3$ and $5$ that approaches zero.
This could be the result of the array being in a different position on the chip which can have a slightly different etch depth.

\end{document}